\documentclass[preprint,10pt,number]{elsarticle}
\usepackage{amsmath,amssymb,booktabs,graphicx,array,hyperref,longtable,url,placeins,float}
\journal{International Journal of Critical Infrastructure Protection}
\begin{document}
\begin{frontmatter}
\title{Closing the Applied Cybersecurity Gap in Critical Infrastructure: A Cyber-to-Physical Security Assessment Framework for Water Systems}
\author[cs]{Halil I. Dursunoglu\corref{cor}}
\ead{halilibrahim.dursunoglu@wmich.edu}
\author[cis]{Kaan Sulkalar}
\ead{kaan.sulkalar@wmich.edu}
\cortext[cor]{Corresponding author}
\affiliation[cs]{organization={Department of Computer Science, Western Michigan University},city={Kalamazoo},state={Michigan},country={USA}}
\affiliation[cis]{organization={Department of Computer Information Systems, Western Michigan University},city={Kalamazoo},state={Michigan},country={USA}}
\begin{abstract}
The integration of information technology and operational technology in water infrastructure creates pathways through which cyber compromise may affect physical service delivery. This study revisits the Applied Cybersecurity Gap (ACG) framework by evaluating an experimental Applied Cybersecurity Gap Index (ACGI) alongside the Cyber-to-Physical Propagation Rate (CPPR), defined as the proportion of initially compromised trials that produce a specified physical-impact outcome. The expanded evaluation comprises 2,160 synthetic trials across three water-network benchmarks, including EPANET Example Network 3 (Net3), and couples declared cyber-attack progression rules with hydraulic simulations. A source-trained machine-learning model evaluated on 247 compromised Net3 trials achieved ROC-AUC 0.772, PR-AUC 0.775, and F1 0.733, compared with 0.661, 0.642, and 0.704 for a scenario-only baseline. The configuration-cluster bootstrap interval for the ROC-AUC difference was 0.038--0.183. A separate 720-run Net3 hydraulic operating-point sensitivity analysis achieved the study's exploratory baseline pressure criterion in all reruns, preserved 717 of 720 physical-impact labels, and produced ROC-AUC 0.791 with the frozen source-trained model. Additional analyses examine the sensitivity of CPPR to the physical-impact threshold and assumed control-blocking probabilities, and demonstrate that the ACGI--CPPR association is partly induced by the simulation design rather than independently validated. The findings establish a reproducible simulation-based assessment workflow, but do not constitute field calibration of cybersecurity-control effectiveness or prediction of real-world utility incidents.
\end{abstract}
\begin{keyword}Operational technology \sep Water distribution \sep Cyber-to-physical propagation \sep Applied Cybersecurity Gap \sep Cross-network transfer \sep WNTR \sep Hydraulic sensitivity
\end{keyword}
\end{frontmatter}
\section{Introduction}
Critical infrastructure systems increasingly depend on the integration of information technology (IT), operational technology (OT), communication networks, and automated control mechanisms to maintain essential services. Water distribution infrastructure represents a particularly important example of this integration because its operational continuity directly affects public health, economic activity, emergency response, and community resilience. Modern water utilities employ supervisory control and data acquisition (SCADA) systems, programmable logic controllers (PLCs), remote monitoring devices, and network-connected actuators to regulate pumping operations, storage levels, pressure conditions, and water delivery. Although these technologies improve operational visibility and efficiency, they also introduce cybersecurity dependencies into physical processes that were traditionally managed within more isolated industrial environments.

The convergence of IT and OT creates security challenges that differ from those encountered in conventional enterprise networks. A cybersecurity incident affecting a business application may primarily compromise information confidentiality, integrity, or availability, whereas an incident involving industrial control systems can also alter physical operating conditions. Unauthorized changes to pump commands, controller settings, sensor measurements, or communication pathways may disrupt the coordination between hydraulic infrastructure and its supervisory systems. Depending on the affected assets, operating conditions, and available safeguards, these disruptions may contribute to abnormal pressure, storage-level deviations, equipment stress, or interruptions in service. Consequently, evaluating cybersecurity in water infrastructure requires consideration of both digital compromise and its potential consequences for physical operations. NIST SP 800-82 Revision 3 emphasizes the distinctive performance, reliability, and safety requirements of OT environments and the importance of incorporating these requirements into cybersecurity practices \cite{nist80082}.

Cybersecurity frameworks provide established mechanisms for identifying assets, assessing threats, implementing safeguards, and organizing risk-management activities. The NIST Cybersecurity Framework 2.0, for example, establishes high-level cybersecurity outcomes that organizations can use to structure governance and operational improvements \cite{nistcsf20}. However, the existence of a cybersecurity policy, architectural requirement, or prescribed control does not necessarily demonstrate that the corresponding protection has been fully implemented or remains effective under operational conditions. Differences may arise from legacy equipment, inconsistent access-control enforcement, incomplete network segmentation, delayed maintenance, limited monitoring coverage, or constraints imposed by continuous industrial operations. These differences are particularly consequential in environments where multiple security mechanisms must function together to prevent an initial intrusion from reaching operational assets.

This study examines such differences through the concept of the \emph{Applied Cybersecurity Gap (ACG)}, defined as the discrepancy between intended cybersecurity protection and its implementation within an operational environment. The ACG concept emphasizes the distinction between documenting a security requirement and establishing that the corresponding safeguard is present and operationally meaningful. To support quantitative experimentation, the study uses an \emph{Applied Cybersecurity Gap Index (ACGI)} that aggregates five normalized dimensions: external exposure, access control, IT/OT segmentation, detection capability, and recovery capability. These dimensions represent different aspects of the protection environment and are modeled as explicit experimental inputs. The index is intended to provide an interpretable representation of security-control implementation gaps rather than replace established vulnerability-scoring systems or serve as an independently validated measure of organizational cyber risk.

A further challenge arises from the distinction between successful initial compromise and the subsequent development of physical consequences. An attacker obtaining access to an exposed digital asset does not necessarily gain control of operational equipment, and reaching an OT component does not necessarily produce a measurable disruption. Network segmentation, access restrictions, controller behavior, process constraints, and hydraulic operating conditions may interrupt or modify the progression of an attack. Conversely, a seemingly limited compromise may have substantial operational implications when it affects an asset with direct influence over the physical process. Treating these stages as a single binary security outcome can obscure the mechanisms through which cybersecurity weaknesses become infrastructure-level consequences.

To examine this progression, the study employs the \emph{Cyber-to-Physical Propagation Rate (CPPR)}, defined as the proportion of initially compromised experimental trials that produce a specified physical-impact outcome. This conditional measure distinguishes the initial compromise event from the subsequent occurrence of measurable physical disruption. In the proposed evaluation, cyber compromise and OT reachability are generated using declared security-dependent simulation rules, while physical-impact outcomes are obtained from hydraulic simulations. The separation of these stages makes the experimental assumptions explicit and permits analysis of how modeled security conditions, attack scenarios, and hydraulic responses contribute to the observed propagation outcomes. CPPR is therefore interpreted as a conditional outcome measure within the simulated environment, not as an empirically established probability of disruption in operational water utilities.

Existing research has provided important foundations for studying cyber-physical security in water infrastructure. Physical and cyber-physical testbeds, including Secure Water Treatment (SWaT) and Water Distribution (WADI), have enabled investigation of attack scenarios, process monitoring, and anomaly detection under controlled experimental conditions \cite{adepu2016,ahmed2017wadi,inoue2017}. The Battle of the Attack Detection Algorithms (BATADAL) has further supported comparative evaluation of detection methods for water distribution networks \cite{taormina2018}. Model-based approaches have investigated the use of hydraulic behavior and residual analysis to identify potentially malicious disturbances \cite{housh2018}. Collectively, these studies demonstrate the importance of incorporating process knowledge into cybersecurity assessment rather than relying exclusively on conventional network-security indicators.

More recent investigations have expanded the application of machine learning, hydraulic models, and digital-twin approaches to cyber-physical water-system security. Parajuli and Shin examined machine-learning-based identification of failure types in cyber-physical water distribution networks \cite{aqua2024}, while Pandey et al. investigated digital-twin-supported anomaly detection and localization \cite{pandey2024}. Raza and Moazeni proposed a framework integrating cyberattack detection, localization, and severity assessment in smart water systems \cite{raza2025}. Homaei et al. explored causal digital twins for anomaly detection in water-related cyber-physical environments \cite{homaei2026}. Recent reviews also document the growing diversity of artificial-intelligence-based attack-detection techniques and the challenges associated with evaluating them across heterogeneous water-system datasets \cite{review2026}. These developments provide a strong methodological foundation, but their principal evaluation targets often concern attack detection, event classification, localization, or severity estimation rather than the conditional prediction of physical impact following an established cyber compromise.

The present research develops an integrated assessment and evaluation framework connecting cybersecurity implementation gaps, simulated attack progression, hydraulic-process responses, and machine-learning prediction. The framework combines explicitly defined security-control variables with six attack-scenario categories and paired baseline-versus-attack hydraulic simulations. It evaluates three network benchmarks, each containing 720 experimental trials. Two synthetic networks, designated A and B, support bidirectional transfer experiments and source-model training, while EPANET Example Network 3 (Net3) provides an additional held-out hydraulic benchmark. Net3 includes 92 junctions, 117 pipes, two pumps, and three storage tanks. Its inclusion enables assessment of physical-impact outcomes in a network with a more detailed hydraulic structure than the simplified representations used in earlier versions of this research.

The contribution of this research is fourfold. First, it formalizes an experimental representation of applied cybersecurity-control gaps and distinguishes this construct from established vulnerability and risk-management measures. Second, it defines a conditional cyber-to-physical propagation outcome that separates initial compromise, OT reachability, and physical-process impact. Third, it implements a reproducible multi-network evaluation protocol combining synthetic security scenarios, WNTR-based hydraulic simulation, source-only machine-learning transfer, feature ablations, and configuration-aware uncertainty estimation. Fourth, it examines the sensitivity of the resulting physical-impact labels and predictive performance to a documented change in the hydraulic operating point. Together, these elements provide an integrated methodology for investigating the relationship between modeled cybersecurity weaknesses and simulated physical consequences in water infrastructure.

\section{Related work and positioning}\label{sec:related}
NIST SP 800-82 Rev. 3 provides control-oriented OT security guidance, not a trial-level probability of physical propagation \cite{nist80082}. The NIST Cybersecurity Framework describes organizational outcomes rather than a water-network-specific conditional propagation estimator \cite{nistcsf20}. SWaT and WADI research investigates attacks and process telemetry in physical testbeds \cite{inoue2017,adepu2016,ahmed2017wadi}. BATADAL compares attack-detection algorithms for water distribution systems \cite{taormina2018}. Our objective differs: predict \emph{conditional physical propagation after a synthetic cyber foothold} and test transfer across different simulated hydraulic networks. We do not claim superior attack detection over those systems because their data, tasks, and experimental protocols differ.

Table~\ref{tab:related} summarizes the scope and limitations of representative water-sector cybersecurity approaches.
\begin{table}[!htbp]
\centering\small
\caption{Representative related approaches and differences in evaluated tasks.}\label{tab:related}
\begin{tabular}{p{.20\linewidth}p{.27\linewidth}p{.27\linewidth}p{.18\linewidth}}\toprule
Approach & Primary method or output & Strength & Not directly assessed \\midrule
NIST OT guidance \cite{nist80082} & OT controls and engineering guidance & Operational control scope & Trial-level CPPR \\
SWaT process invariants \cite{adepu2016} & Physical-process anomaly rules & Process-aware attack evidence & Cross-network CPPR transfer \\
WADI \cite{ahmed2017wadi} & Physical water-distribution testbed & Physical realism & This paper's ACGI target \\
BATADAL \cite{taormina2018} & Comparative attack detection & Common attack-detection benchmark & Conditional post-compromise outcome \\
Model-based hydraulics \cite{housh2018} & Hydraulic residuals for attack detection & Mechanistic operational baseline & Conditional post-compromise risk \\
Stealthy hydraulic attacks \cite{stealth2025} & False-data-injection attack impacts & Attack-impact mechanisms & Our security-gap index \\
Digital twin localization \cite{pandey2024} & Learning-based anomaly localization & Network-model integration & Our target-network transfer \\
This study & ACGI, CPPR, synthetic hydraulics and source-only transfer & Explicit held-out-network and operating-point sensitivity & Field validation and fair same-task SOTA reproduction \\\bottomrule
\end{tabular}
\end{table}

\subsection{Recent reviewer-identified studies and novelty boundary}
A systematic review by Tuptuk et al.~\cite{tuptuk2021} surveys the cybersecurity landscape of water systems, including vulnerabilities, attack surfaces, and challenges in safeguarding interconnected operations. This sector-wide synthesis motivates assessment of implemented controls but does not supply a calibrated conditional post-compromise propagation metric. Raza and Moazeni~\cite{raza2025} developed optimized reconstruction-based attack detectors, localization methods, and a severity index for water-distribution networks using BATADAL and C-Town. Their severity construct quantifies attack consequences and must be acknowledged as prior work on quantitative water-system impact assessment; our CPPR instead defines a binary physical-impact frequency conditional on an initial foothold, and is not claimed to be the first impact measure.

Wang et al.~\cite{wang2026} developed output-constrained secure second-order sliding-mode control for drinking-water coagulation under false-data-injection and denial-of-service attacks, with theoretical stability analysis and experimental case studies. Their work addresses preservation of physical-process control performance under attack rather than estimation of security-gap scores. Voggensberger et al.~\cite{voggensberger2027} introduced a context-aware methodology for selecting and weighting cybersecurity controls for water-sector industrial control systems. Their study is particularly relevant to ACGI: context-dependent and weighted control assessment is established prior art, and the present work does not claim to originate either concept. Its distinct methodological question is whether declared control-gap assumptions can be linked to a separately defined conditional propagation endpoint and tested for consistency under hydraulic operating changes and source-to-target network transfer. Because the security assumptions remain synthetic, this linkage is not independent field validation of control effectiveness.

Table~\ref{tab:reviewerrecommended} compares the four reviewer-recommended studies with the present framework and identifies the limits of direct performance comparisons.
\begin{table}[!htbp]\centering\footnotesize
\caption{Direct comparison with the four publications identified by Reviewer 1. Reported performance across these studies is not numerically comparable because outcomes and experimental protocols differ.}\label{tab:reviewerrecommended}
\begin{tabular}{p{.21\linewidth}p{.26\linewidth}p{.27\linewidth}p{.18\linewidth}}\toprule
Study & Primary contribution & Evaluation basis & Distinction from this work \\ \midrule
Tuptuk et al.~\cite{tuptuk2021} & Systematic review of water cybersecurity & Published research synthesis & Not a trial-level propagation experiment \\
Raza and Moazeni~\cite{raza2025} & Attack detection, localization, severity index & BATADAL and C-Town & Severity/detection rather than conditional CPPR \\
Wang et al.~\cite{wang2026} & Output-constrained secure coagulation control & Analytical, simulated and experimental control evaluation & Resilient control rather than control-gap measurement \\
Voggensberger et al.~\cite{voggensberger2027} & Context-aware selection and weighting of ICS security controls & Plant-context security assessment methodology & Strong overlap with control weighting; no claimed novelty for weighted scoring \\
This study & Synthetic ACGI, conditional CPPR, cross-network prediction & A/B/Net3 simulation and WNTR sensitivity & No empirical security calibration or expert interviews \\ \bottomrule
\end{tabular}
\end{table}

Recent reviews of security in water treatment and distribution summarize threat classes, industrial control dependencies, and the practical constraints of intrusion detection \cite{gulzar2024}. For instance, the BATADAL benchmark compares attack detection on a common water-network benchmark \cite{taormina2018}, while model-based fault detection uses hydraulic state estimation to identify cyber-induced anomalies \cite{housh2018}. These are useful methodological precedents, but they do not directly estimate the post-compromise physical-impact endpoint used here. More recent research has investigated stealthy false-data-injection effects on hydraulic operations \cite{stealth2025}, flood consequences of attacks on smart stormwater controls \cite{lin2024}, and machine-learning digital twins for distribution-network anomaly localization \cite{pandey2024}. The distinctions among detection, localization, physical severity and conditional propagation motivate reporting explicit target definitions rather than comparing unlike accuracy scores.\par

Recent work has also studied failure classification with WNTR and epanetCPA \cite{aqua2024}, temporal graph models on BATADAL \cite{tgn2021}, and a 2025 hybrid classifier for water-system attack detection \cite{chandran2025}. These methods target event detection or failure identification, whereas the present task is prediction of a physical-impact label conditional on a simulated initial foothold. Their reported scores cannot be fairly compared numerically without reconstructing a common outcome definition, observation window, and train/test protocol. A recent systematic review synthesizes AI-based water-distribution attack-detection approaches and highlights the variety of benchmark settings \cite{review2026}. Raza and Moazeni combine detection, localization, and a severity index in C-Town \cite{raza2025}; Homaei et al. investigate causal digital twins and counterfactual reasoning on SWaT, WADI and HAI \cite{homaei2026}. These studies sharpen the distinction between event detection, impact severity, and the present conditional propagation target. Neither their published detection scores nor their testbed results are directly comparable with the synthetic post-compromise labels used here.\par

For hydraulic-model reproducibility, the EPANET 2.2 manual describes the extended-period network state variables and solution assumptions \cite{rossman2020}; the WNTR manual and its associated software-framework publication document Python-based resilience experiments \cite{wntr2023,klise2017}. Broader water-network research on operational hydraulic anomalies and energy management \cite{duan2025} further emphasizes that hydraulic feasibility and security-event prediction are related but distinct evaluation problems.\par

\section{Materials and methods}\label{sec:methods}
\subsection{Research workflow and study design}
\begin{figure}[!htbp]\centering\includegraphics[width=\linewidth]{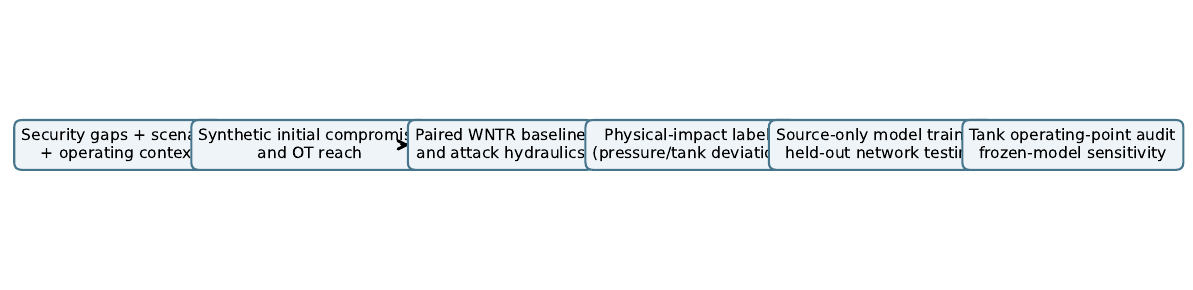}\caption{Study workflow. Security outcomes are synthetic; physical labels come from paired hydraulic trajectories; target-network fitting is prohibited.}\label{fig:workflow}\end{figure}

The workflow is: (i) generate a security configuration and attack scenario; (ii) simulate initial compromise and OT reach using declared rules; (iii) run paired baseline and attack hydraulic trajectories; (iv) label material pressure or tank deviation; (v) train on source networks only; (vi) evaluate on held-out networks; (vii) audit baseline pressure and conduct a separate Net3 operating-point sensitivity. No attack-induced hydraulic quantity or target label is included in pre-attack model features. We report precision--recall AUC alongside ROC-AUC because the prevalence of physical-impact labels is not fixed across networks, and PR evaluation can be especially informative under class imbalance \cite{saito2015}. Machine-learning implementation follows established Python scientific-learning conventions \cite{pedregosa2011}.
The sequence of security simulation, paired hydraulic evaluation, and held-out-network prediction is illustrated in Fig.~\ref{fig:workflow}.

\subsection{Security-gap construct and conditional propagation}
Let $g=(E,A,S,D,R)$ denote five normalized implementation gaps: external exposure, access, IT/OT segmentation, detection, and recovery, each in $[0,1]$ with higher values representing weaker protection. We use
\begin{equation}\mathrm{ACGI}_{w}=\sum_{j=1}^{5}w_jg_j,\qquad w_j\geq0,\quad\sum_jw_j=1.\end{equation}
Equal weights ($w_j=0.2$) form an interpretable reference, not empirically estimated causal weights. ACGI is an experimental composite of predefined gap inputs, not a substitute for CVSS, risk-based asset prioritization, or an externally validated security score. In the primary combined predictor, component gaps are used directly rather than simultaneously including their deterministic weighted sum. Earlier A/B ablations included alternative weighting and leave-one-dimension-out tests; they are sensitivity analyses, not evidence that a universal set of weights has been identified.

For $I_i$ indicating initial compromise and $Y_i$ indicating physical impact, define
\begin{equation}\widehat{\mathrm{CPPR}}=\frac{\sum_i\mathbf{1}(I_i=1,Y_i=1)}{\sum_i\mathbf{1}(I_i=1)}.\end{equation}
This is an empirical conditional fraction in the \emph{synthetic experiment}. For example, Net3 produced 247 initial compromises and 137 impacts, so its empirical CPPR is $137/247=0.555$. A different denominator, 172 trials that reached OT, gives $137/172=0.797$ and must not be called the same quantity.

\subsection{Hydraulic networks, dataset provenance, and scenarios}
The study uses two synthetic water-network configurations (A and B) and EPANET Example Network 3 (Net3) loaded from the WNTR distribution. Each network has 720 generated trials, giving 2,160 trials before filtering. A and B yielded 217 and 185 compromised trials, respectively; Net3 yielded 247. Net3 includes 92 junctions, 117 pipes, two pumps, three tanks, and no physical valves. The input network file was tracked using a SHA-256 digest recorded in the reproducibility manifest. These are \emph{newly generated trial-level datasets}, not field-collected attack incidents.

The six attack labels are remote access, PLC setpoint manipulation, sensor spoofing, valve manipulation, pump override, and command replay. In Net3, the valve-manipulation label is implemented by isolating pipe 205 because Net3 has no valve; it is explicitly a \emph{pipe-isolation proxy}. Other scenarios share simplified pump-status interventions, while replay alternates status. Scenario names should not be read as six independently validated physical attack mechanisms. The Net3 design includes 120 trials per scenario (six scenarios, 720 trials), distributed over 40 security-configuration clusters. The generator determines initial compromise and OT reach before hydraulics; this distinction is essential because physical outcomes are never independent of those rules. For a concrete interpretation, a configuration with component gaps $(0.8,0.6,0.4,0.2,0.0)$ has equal-weight ACGI $0.40$; this arithmetic example is illustrative rather than an observed utility measurement. ACGI alone cannot identify which control should be remediated without examining the five components and attack paths. Each attack trajectory is compared against a matched no-attack trajectory with the same operating inputs. A physical impact is flagged for OT-reached trials when the maximum pressure deviation reaches 5 m or the maximum tank-level deviation reaches 0.35 m. These are exploratory research thresholds, not statutory water-service limits.

Hydraulics are solved using \texttt{WNTRSimulator} in demand-driven mode. Baselines and attacked runs are matched; pressure is evaluated at five-minute reporting timestamps. Net3 junction 10, an immediate pump-adjacent zero-demand node, is excluded from the service-pressure feasibility screen after a documented audit found localized negative pressure there. Its exclusion is not used to erase or change physical-impact labels: an earlier all-versus-excluding-junction-10 rerun changed zero labels in 720 trials. The 91 other junctions remain monitored. The study does not establish that all nodes satisfy regulatory pressure requirements.

\subsection{Prediction, ablations, and leakage safeguards}
The prediction target is $Y$ \emph{conditional on} $I=1$. The source-only training set consists of compromised trials from A and B ($217+185=402$); Net3 is held out (247 compromised trials). Additional tests train on A and evaluate on B, and vice versa. Four model inputs are compared: combined security gaps and scenario/operating features; scenario-only; security-only; and hydraulic-only. The original experimental implementation is deterministically reconstructed and checked against saved predictions; maximum absolute prediction discrepancy is below $10^{-6}$. Net3 labels, features, and threshold are never used to refit or tune the source models.

Trials share a configuration and therefore are not independent observations. Holdout is by network, and uncertainty is estimated by resampling target configuration IDs (3,000 resamples). This captures target-configuration sampling variation conditional on a fitted source model; it does not capture model-retraining uncertainty, parameterization uncertainty, or field deployment variation. Performance measures are ROC-AUC, PR-AUC, F1 at the preselected 0.5 threshold, and Brier score. AUC comparisons are made against matched ablations; they are not presented as head-to-head results against unrelated published detection systems.

\subsection{Hydraulic operating-point sensitivity}
The original Net3 operating point was infeasible under an exploratory minimum pressure of 5 m at the 91 service-monitored junctions. A 24-trial demand-factor pilot tested 264 baseline evaluations and found no feasible factor; junction 40 was consistently limiting. A separate predefined screen of allowable initial tank-level increments selected a +4 m adjustment to Tank 1, within its stated tank-level bounds. This selection used baseline pressure only, not attack labels. The complete 720-trial experiment then reran matched hydraulics with all cyber outcomes and impact thresholds fixed. A further source-only ML sensitivity increased the mean initial tank-level feature by $4/3$ m (three tanks) and evaluated adjusted labels without target retraining.

\section{Results}\label{sec:results}
\subsection{Cross-network transfer}
\begin{figure}[!htbp]\centering\includegraphics[width=.85\linewidth]{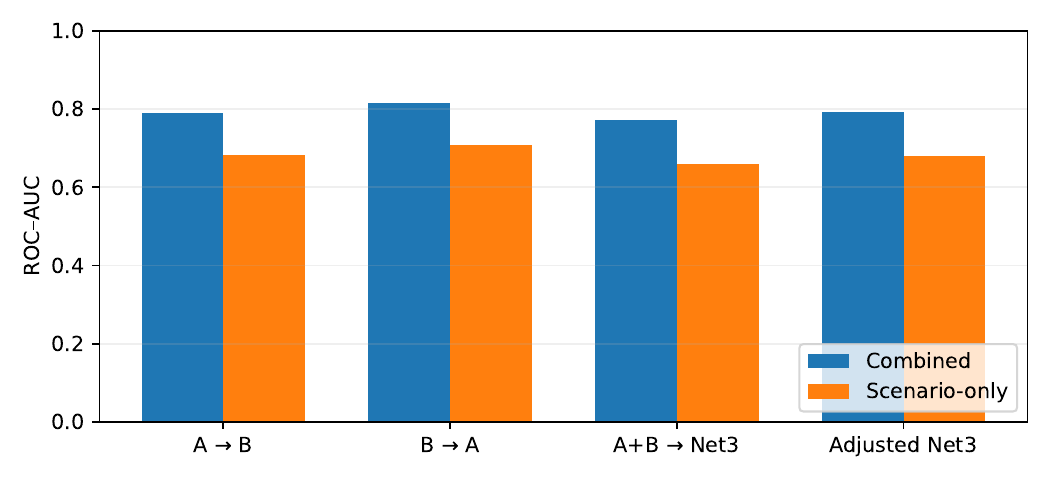}\caption{Held-out-network discrimination for combined and scenario-only models. The adjusted Net3 evaluation uses revised physical-impact labels.}\label{fig:auc}\end{figure}

Table~\ref{tab:transfer} reports cross-network predictive performance on initially compromised trials, including the held-out Net3 evaluation and scenario-only baselines.
\begin{table}[!htbp]\centering\small
\caption{Completed transfer evaluation on compromised trials. F1 uses a fixed 0.5 cutoff.}\label{tab:transfer}
\begin{tabular}{llrrrr}\toprule
Train $\to$ test & Model & $n$ & ROC-AUC & PR-AUC & F1 \\midrule
A $\to$ B & Combined & 185 & .791 & .770 & .688 \\
A $\to$ B & Scenario-only & 185 & .681 & .578 & .689 \\
B $\to$ A & Combined & 217 & .815 & .780 & .735 \\
B $\to$ A & Scenario-only & 217 & .707 & .622 & .746 \\
A+B $\to$ Net3 & Combined & 247 & .772 & .775 & .733 \\
A+B $\to$ Net3 & Scenario-only & 247 & .661 & .642 & .704 \\
A+B $\to$ Net3 & Security-only & 247 & .670 & .686 & .696 \\
A+B $\to$ Net3 & Hydraulic-only & 247 & .522 & .581 & .714 \\\bottomrule
\end{tabular}
\end{table}
The held-out discrimination comparisons, including the adjusted Net3 analysis, are visualized in Fig.~\ref{fig:auc}.
The combined model improves Net3 ROC-AUC over scenario-only by 0.112 (95\% configuration-cluster bootstrap CI 0.038--0.183) and PR-AUC by 0.134 (CI 0.067--0.206). Its F1 improvement is 0.029, but the bootstrap interval ($-0.087$ to 0.121) crosses zero. A $\to$ B and B $\to$ A exhibit ROC-AUC improvements of approximately 0.109 and 0.108, respectively. These outcomes address RQ1 and RQ2 within the synthetic setting, while the lack of F1 separation cautions against universal superiority claims.

\subsection{Held-out Net3 confusion matrix}
The held-out Net3 confusion matrix is presented in
Table~\ref{tab:cm}. Among the 247 initially compromised
trials, the combined model correctly identified 99 of
137 physical-impact cases and 76 of 110 non-impact
cases. It produced 34 false positives and 38 false
negatives, corresponding to a precision of 0.744,
recall of 0.723, and F1-score of 0.733.
\begin{table}[!htbp]
\centering
\caption{Confusion matrix for the combined model on the
held-out Net3 dataset, conditional on initial compromise
($n=247$).}
\label{tab:cm}
\begin{tabular}{lrrr}
\toprule
\textbf{Actual class} &
\multicolumn{2}{c}{\textbf{Predicted class}} &
\textbf{Total} \\
\cmidrule(lr){2-3}
& Negative & Positive & \\
\midrule
Negative & 76 & 34 & 110 \\
Positive & 38 & 99 & 137 \\
\midrule
Total & 114 & 133 & 247 \\
\bottomrule
\end{tabular}
\end{table}

\subsection{Security-gap weighting sensitivity}
The ACGI weighting experiment evaluates equal weighting, five single-dimension-dominant schemes (weight 0.5 on one gap and 0.125 on each remaining gap), and five leave-one-dimension-out schemes (weight zero on one gap and 0.25 on each remaining gap). The following values summarize the mean ROC--AUC over the five saved source-network folds; they are not a held-out Net3 comparison, and fold means are descriptive rather than independent confidence intervals.
Table~\ref{tab:weights} reports the descriptive source-fold ROC--AUC results under alternative gap-weighting schemes.
\begin{table}[!htbp]\centering\small\caption{ACGI weighting sensitivity: mean ROC--AUC across five source folds.}\label{tab:weights}
\begin{tabular}{lr}\toprule Weight scheme & Mean ROC--AUC \\ \midrule
segmentation gap dominant & 0.801 \\
without access gap & 0.746 \\
without exposure gap & 0.731 \\
without detection gap & 0.727 \\
equal & 0.707 \\
without recovery gap & 0.694 \\
recovery gap dominant & 0.687 \\
access gap dominant & 0.673 \\
without segmentation gap & 0.671 \\
exposure gap dominant & 0.656 \\
detection gap dominant & 0.655 \ \\\bottomrule\end{tabular}\end{table}
The weighting exercise is a robustness probe of a constructed index; it cannot establish that any weight vector is an unbiased measure of real-world protection. For deployment, control criticality and context would require independent calibration.

\subsection{Hydraulic feasibility and label stability}
\begin{figure}[!htbp]\centering\includegraphics[width=.86\linewidth]{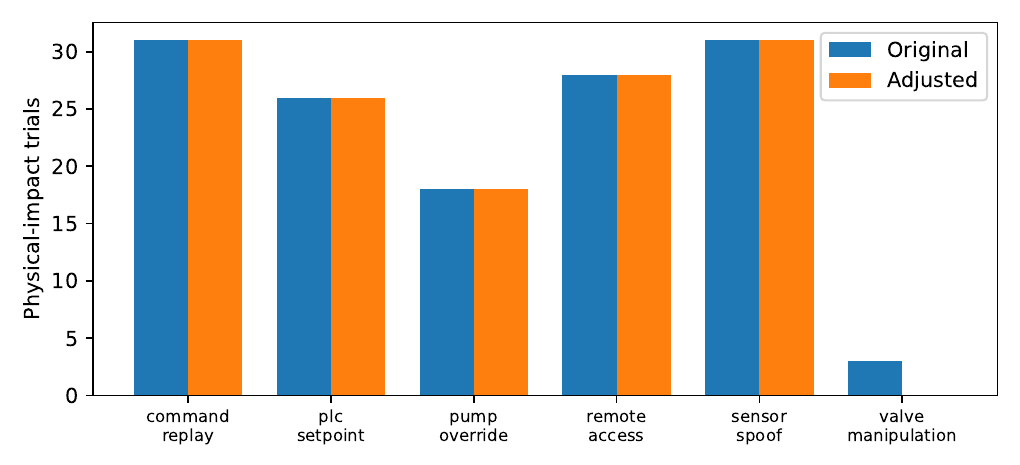}\caption{Net3 scenario-level physical-impact labels before and after the operating-point change (120 trials per scenario). The three changes are confined to the pipe-isolation proxy.}\label{fig:labels}\end{figure}

At the original operating point, all 720 Net3 baselines had a monitored service-pressure minimum below 5 m. Excluding junction 10 alone did not change any of 720 impact labels, but did not repair that pressure shortfall. A demand-reduction pilot found no feasible factor in 264 evaluations. An initial Tank 1 level increase of 4 m yielded baseline service-pressure feasibility in 720/720 reruns, with zero recorded tank-bound violations. Positive labels decreased from 137 to 134: 717/720 (99.58\%) remained unchanged. The three positive-to-negative changes all occurred in the pipe-isolation proxy; the other five scenarios preserved their labels. Configuration-bootstrap disagreement-rate 95\% CI was 0--0.97\%. Figure~\ref{fig:labels} shows the scenario-level distribution of labels before and after the operating-point adjustment. 
This addresses RQ3 only for the declared 5 m, five-minute-sampled, 91-junction feasibility criterion.

\subsection{Frozen-model operating-point sensitivity}
Table~\ref{tab:adjust} compares the frozen source-trained models at the original and adjusted Net3 operating points.
\begin{table}[!htbp]\centering\small
\caption{Original and adjusted Net3 transfer; source-only models, no target tuning.}\label{tab:adjust}
\begin{tabular}{llrrrr}\toprule
Operating point & Model & ROC-AUC & PR-AUC & F1 & Brier \\midrule
Original & Combined & .772 & .775 & .733 & .191 \\
Adjusted & Combined & .791 & .779 & .744 & .179 \\
Adjusted & Scenario-only & .680 & .643 & .711 & .198 \\
Adjusted & Security-only & .666 & .672 & .689 & .233 \\
Adjusted & Hydraulic-only & .508 & .561 & .703 & .249 \\\bottomrule\end{tabular}
\end{table}
Adjusted combined versus scenario-only ROC-AUC difference is 0.111 (95\% CI 0.036--0.182), and PR-AUC difference is 0.136 (CI 0.067--0.210). Changing only the mean tank-level input while retaining original labels leaves ROC-AUC and PR-AUC unchanged; the adjusted-label evaluation accounts for the observed discrimination difference. We therefore do not attribute the AUC improvement to tank adjustment alone.

\subsection{Trial provenance and reproducibility controls}
The Stage 5.7 manifest records SHA-256 hashes for all three source/target trial tables, the adjusted hydraulic trial table, and the original reference predictions. The source-only reconstruction matches the original four models to floating-point precision (maximum absolute difference $1.11\times 10^{-16}$). In the sensitivity experiment, the only changed pre-attack hydraulic feature is mean initial tank level, increased by $4/3$ m; security gaps, scenario identities, and other pre-attack features are unchanged. The 3,000 configuration-cluster bootstrap replicates are conditional on the reconstructed models. Data-generation code, trial CSVs, configuration IDs, model reconstruction and operating-point manifests should accompany the submission in a versioned archive.

\subsection{Original-testbed parameter provenance, decoupling, and sensitivity}
\label{sec:reviewer2}
The originally submitted 720-trial single-tank experiment is retained as a methodological diagnostic rather than independent validation. Inspection of the released simulator's \texttt{security.py} and \texttt{experiment.py} shows that five fixed gap scores define the ACGI, whereas distinct, configuration-assigned Bernoulli blocking probabilities determine remote-access, lateral-movement, controller-access, and controller-integrity outcomes. The gap scores are \emph{not} directly converted to blocking probabilities by an estimated function. Instead, both are assigned together to four cumulative, deliberately ordered configurations. These values were illustrative design assumptions, not elicited from operators, measured at utilities, or calibrated from incident data. Consequently, an observed ACGI--CPPR correlation cannot be presented as independent predictive or causal validation.

The complete configuration-specific security-stage assumptions are reported in Table~\ref{tab:r2parameters}; these values are simulation inputs, not field-calibrated estimates.
\begin{table}[!htbp]\centering\scriptsize
\caption{Original testbed's complete assumed security-stage parameters; probabilities are configuration inputs, not empirical estimates.}\label{tab:r2parameters}
\resizebox{\linewidth}{!}{%
\begin{tabular}{lrrrrrrr}\toprule
Config. & Auth block & Lateral block & Controller block & Integrity block & Recovery eff. & Cyber det. & Process det.\\\midrule
C0 & .10 & .10 & .08 & .05 & .25 & .30 & .20\\
C1 & .68 & .18 & .15 & .10 & .35 & .50 & .30\\
C2 & .76 & .78 & .62 & .30 & .50 & .64 & .42\\
C3 & .84 & .88 & .82 & .82 & .82 & .78 & .90\\\bottomrule
\end{tabular}}
\end{table}

For each attack scenario, the \texttt{needs\_remote\_access}, \texttt{needs\_lateral\_movement}, and \texttt{needs\_controller\_access} Boolean flags determine which Bernoulli blocking gates apply in sequence. A failed remote-access gate prevents initial compromise; a failed lateral gate prevents OT reach; a failed controller gate prevents controller access. The integrity gate applies only to persistent threshold modification. Detection can probabilistically trigger containment through \texttt{recovery\_efficiency}; the process-detection strength is used in scoring, not as a direct independent hydraulic failure probability. The original binary physical-impact criterion was controller access and a maximum attacked-versus-counterfactual \emph{normalized tank-level} difference of at least 0.08. The 0.08 value is a dimensionless exploratory cutoff, not a physical pressure standard or field-calibrated safety threshold. This single-tank criterion differs from the later Net3 pressure/tank-level thresholds, which must not be conflated.

Table~\ref{tab:r2threshold} presents the recomputed CPPR under alternative normalized physical-impact thresholds.
\begin{table}[!htbp]\centering\small
\caption{Recomputed CPPR from the original 720 saved trial trajectories at alternative normalized tank-level impact thresholds. Denominators remain 176, 161, 161, and 157 initial compromises for C0--C3.}\label{tab:r2threshold}
\begin{tabular}{lrrrr}\toprule
Threshold & C0 & C1 & C2 & C3\\\midrule
0.04 & .864 & .640 & .118 & .051\\
0.06 & .847 & .615 & .118 & .051\\
0.08 (original) & .784 & .540 & .118 & .051\\
0.10 & .761 & .497 & .106 & .045\\
0.12 & .744 & .484 & .081 & .032\\\bottomrule
\end{tabular}
\end{table}

The threshold sweep reclassifies \emph{saved} trajectories without changing the process or attack-generation mechanism. Reclassification at 0.08 reproduces all original labels. The rank order C0$>$C1$>$C2$>$C3 persists throughout the explored threshold range, but the absolute rates are threshold-dependent.

Table~\ref{tab:r2blocking} reports the effects of multiplicative perturbations to the original simulator's stage-blocking probabilities.
\begin{table}[!htbp]\centering\small
\caption{Rerun CPPR after multiplicative perturbation of the four stage-blocking probabilities in the original simulator (common trial seeds; probabilities clipped to [0,1]). Recovery and detector settings are held fixed.}\label{tab:r2blocking}
\begin{tabular}{lrrrr}\toprule
Blocking-probability multiplier & C0 & C1 & C2 & C3\\\midrule
0.8 & .795 & .617 & .220 & .110\\
1.0 (original) & .784 & .540 & .118 & .051\\
1.2 & .778 & .503 & .065 & .007\\\bottomrule
\end{tabular}
\end{table}

This is a local sensitivity analysis, not an uncertainty distribution for real-world controls. Each multiplier was applied to \texttt{auth\_block\_prob}, \texttt{lateral\_block\_prob}, \texttt{controller\_block\_prob}, and \texttt{integrity\_block\_prob} in every configuration, with original trial seeds and all other settings unchanged. Because branching consumes random draws conditionally, reruns preserve seeds but do not guarantee identical subsequent random-number streams. Rankings remain ordered in these three runs, although the C3 rate varies substantially.

\paragraph{Weighting and decoupling diagnostic.} We additionally sampled 10,000 nonnegative ACGI weight vectors from a five-dimensional Dirichlet(1) distribution (seed 2026) and examined five leave-one-dimension-out averages. Every sampled weight vector preserves the original configuration ordering, because \emph{each of the five assigned gap dimensions} decreases monotonically from C0 through C3. This apparent weight robustness is therefore a mathematical property of the designed configuration table, \emph{not} empirical validation of the weights. To expose this dependence, we also assigned 10,000 independently sampled five-component gap-score vectors to the four fixed outcome configurations without altering their blocking probabilities. The mean Spearman correlation between the independently assigned index and the four original CPPR values was approximately $-0.012$, compared with $1.0$ for the original ordered assignments. This is a negative-control diagnostic: randomly assigned gaps have no reason to predict cyber outcomes. It does not establish that random gaps are a realistic assessment instrument or that any decoupled index has been validated. A valid external ACGI calibration requires measured control evidence independently collected from the attack-outcome generation process.

\section{Discussion and threats to validity}\label{sec:discussion}
\paragraph{Interpretation.} The three-network experiments provide executed evidence that pre-attack security gaps and scenario context can discriminate a synthetically generated physical-propagation outcome on held-out hydraulic networks. The combined model's positive AUC advantage persists under the adjusted Net3 operating point. CPPR is conditional on initial compromise; it is not the probability of attack initiation or an unconditional estimate of utility failure. ACGI provides a transparent summary of specified controls, but its equal weights are design assumptions rather than empirically calibrated contributions.

\paragraph{Circularity and construct validity.} Initial compromise and OT reach are explicitly driven by simulated security-gap rules. A predictor receiving those gaps can learn part of the outcome-generating mechanism. Consequently, high transfer AUC is evidence of consistency across these synthetic environments, not independent confirmation of the causal value of the controls. Likewise, a process-aware detector evaluated against labels defined by process deviation can be circular. For this reason, prior detector precision/recall figures are \emph{not} promoted as an independently validated detection contribution in this revision; any detector study would require independently labeled events and a separate detection benchmark.

\paragraph{Hydraulic validity.} Net3 supplies actual pipe/pump/tank topology rather than a single normalized storage equation, addressing a major earlier omission. Nevertheless, WNTR demand-driven simulation is not a field instrument; the 5 m threshold is exploratory; discrete reports can miss intermediate extrema; the pump-adjacent junction 10 is excluded from the service-pressure criterion; and some attack labels use pump-status or pipe-isolation proxies. The chosen +4 m initial level is a bounded operating-point sensitivity, not a validated operational intervention for a real utility.

\paragraph{Expert assessment and operational applicability.} The framework's five gap dimensions and their equal-weight index have not been independently elicited or validated through structured interviews with utility operators, OT engineers, or water-sector cybersecurity specialists. The experiments therefore cannot establish practitioner agreement on dimension selection, weighting, the realism of attack-stage probabilities, or deployment feasibility. Expert interviews or a documented structured elicitation study are a distinct validation requirement, not a substitute for hydraulic simulation. Such an assessment should recruit participants with complementary water-operations and OT-security expertise, use a consistent protocol to evaluate control observability and scoring, and report disagreement and sensitivity of ACGI rankings to elicited weights. No expert interviews are claimed in the present study.

\paragraph{External validation and benchmarking.} A, B, and Net3 are different simulated hydraulic networks, not three independent real utilities. The study does not implement recent competing cyber-to-physical predictors on an identical data-generating protocol; its fair same-task comparisons are limited to internal ablations. External testing on measured SCADA telemetry, a physical or hardware-in-the-loop water testbed, independently specified cyber compromise events, and broader hydraulic solvers is required before deployment or SOTA claims. Bootstrap intervals resample 40 Net3 configurations conditional on a fitted model and should not be interpreted as uncertainty over utilities.

\section{Conclusion}\label{sec:conclusion}
We presented an experimental framework linking the Applied Cybersecurity Gap Index (ACGI) to the Cyber-to-Physical Propagation Rate (CPPR), defined conditionally on initial cyber compromise. The evaluation comprised 2,160 synthetic trials across three networks. On held-out Net3, 137 of 247 initially compromised trials produced physical-impact labels (CPPR 55.5\%). A source-trained predictor achieved ROC-AUC 0.772 and PR-AUC 0.775, versus 0.661 and 0.642 for a scenario-only comparator; the configuration-bootstrap interval for the ROC-AUC difference was 0.038--0.183. A separate 720-run Net3 operating-point test satisfied the exploratory baseline pressure screen, preserved 717 impact labels, and yielded ROC-AUC 0.791 for the frozen predictor.

The sensitivity and decoupling analyses demonstrate that the observed ACGI--CPPR relationship is partly imposed by the security-dependent simulator, and that propagation estimates depend on the assumed blocking probabilities and physical-impact threshold. These results support a reproducible \emph{simulation-based} assessment protocol, not a field-calibrated risk score, causal control-effectiveness estimate, or real-utility failure forecast. Future validation requires independent cyber events, utility or hardware-in-the-loop measurements, practitioner assessment, and broader hydraulic operating conditions.

\section*{Funding}
This research received no external funding.
\section*{Declaration of competing interest}
The authors declare no competing interests.
\section*{CRediT authorship contribution statement}
\textbf{Halil I. Dursunoglu:} Conceptualization, Methodology, Software, Validation, Formal analysis, Investigation, Data curation, Writing--original draft, Writing--review \& editing, Visualization, Project administration.

\textbf{Kaan Sulkalar:} Conceptualization, Validation, Resources, Writing--review \& editing.

\section*{Ethics statement}
This study used an isolated software-based cyber-physical simulation and did not involve human participants, animals, or interaction with production critical-infrastructure systems. Institutional review board approval was therefore not required.

\section*{Data and code availability} Trial CSVs, simulator scripts, configuration manifests, and frozen-model verification outputs were generated for this study. The associated research archive is identified by DOI \url{https://doi.org/10.5281/zenodo.23269487}. Authors should verify that the deposited release includes the files needed to reproduce each reported analysis.

\end{document}